\documentstyle[amssymb,12pt]{article}

\input{tcilatex}

\begin{document}

\bigskip

\begin{center}
{\huge {\bf The cosmological constant\\[0pt]
as a residual energy in the\\[0pt]
chaotic inflationary model}}\\[0pt]
\bigskip \bigskip

{\Large {\bf De-Hai Zhang}}\\[0pt]
(e-mail: zhang.dh@263.net)\\[0pt]

Department of Physics, Graduate School in Beijing,\\[0pt]
University of Science and Technology of China,\\[0pt]
P.O.Box 3908, Beijing 100039, P.R.China.\\[0pt]

\bigskip

{\bf Abstract:}
\end{center}

{\hspace*{5mm} A new idea of the cosmological constant is proposed in this
paper. Due to the horizon is limited, the quantum fluctuation of the
inflaton field is not zero, a nonzero vacuum energy is remained as a
residual inflationary energy of an unusual potential, however the true
stable vacuum energy is zero fortunately. A unified model of the
cosmological constant and the chaotic inflation is proposed, which satisfies
almost all cosmological phenomenology and will can be tested by data of the
cosmic large scale structure. }

\bigskip

\section{1.Introduction}

{\hspace*{5mm} An astonished result of the high redshift type Ia supernovae
observations in 1998 is that the deceleration parameter comes out negative$%
^{[1,2]}$, implying that our universe is speeding up. The current analyses
favor $\Omega _{m}\simeq 0.4$ and $\Omega _{\Lambda }\simeq 0.6$, and the
universe is flat. The accelerated expansion of the universe may be induced
by a nonzero cosmological constant, which corresponds to $\Lambda _{0}=$ $%
3M_{p}^{2}H_{0}^{2}\ \Omega _{\Lambda }\simeq $ $(2$meV$)^{4}$, where $%
\tilde{M}_{p}=\sqrt{8\pi }M_{p}\simeq 1.2\times 10^{19}$GeV is the Planck
energy scale, and the present Hubble constant is $H_{0}=100h$ km/sec/Mpc $%
\simeq $ $5.6\times $ $10^{-61}M_{p}$ if we take $h\simeq 0.65$. }

The relatively tiny value for a cosmologically relevant $\Lambda _{0}$ seems
very fine tuned from the particle physics point of view and so the prejudice
in a large part of \ the particle physics community has been that it should
be zero, by some mechanism not yet understood$^{[3-5]}$. An extremely small
but nonzero $\Lambda _{0}$ could perhaps be taken as some small effect
perturbing around the preferred value of zero. A lot of new ideas have been
suggested, such as rolling scalar field$^{[6]}$, variable cosmological
constant, X-matter, generalized dark matter, loitering universe,
quintessence model, tracker field and attractor solutions, tangled strings$%
^{[7]}$ and textures$^{[8]}$, and so on. However, one should still continue
to seek a better explanation on the essence of the nonzero and tiny
cosmological constant. The purpose of this letter is to suggest a new model
in which the chaotic inflation and the cosmological constant are unified. In
spite of the true theoretical cosmological constant is zero, the observing
cosmological constant as a relic energy of the inflation potential is
nonzero actually due to the finite horizon and the quantum fluctuation of
the inflaton field.

\section{2.The universality of the quantum fluctuation of the inflaton field}

{\hspace*{5mm} We note that the cosmological constant, i.e., vacuum energy,
has been playing an important role in the inflationary epoch of the early
universe. This unstable vacuum energy is very large during that time. Our
first question is whether the large cosmological constant of the inflation
is same with the present tiny cosmological constant in essence? We think so
and suppose that both can be unified in a common inflationary potential,
i.e., the essence of both is same. The second question is what kind of
potentials should be adopted by us? We suppose that the lowest stable vacuum
must site at a point of $\phi =0$ and $V(\phi )=0$, i.e., the true vacuum
must have a zero cosmological constant from the view point of the
naturalness. The reason that the position of the lowest stable vacuum is
taken at $\phi =0$ by us is that we worry that a nonzero large average value
of the true vacuum $<\phi >\neq 0$ may give easily some particles with
dangerous large masses due to the singlet property of the inflaton $\phi $.
In according to an inference of the inflation theory, after the slow rolling
of the inflation the universe will rapidly roll down to its true vacuum
state, i.e., $V(0)=0$ point, thus there is not any nonzero cosmological
constant to be remained. However, this inference may be wrong. Since many
reasons hint that the inflation field $\phi $ may have a nonzero uncertainty
value $\delta \phi $, thus the actual vacuum energy $V(\delta \phi )$ will
be nonzero. }

In fact the inflaton field $\phi $ disassemble into two parts during its
evolutionary process, the unstable vacuum average value $<\phi >$ and the
quantum fluctuation $\delta \phi $, i.e., $\phi =<\phi >$ $+\delta \phi $.
The quantum fluctuation of de-Sitter universe is determined by its expansion
rate $H=\dot{R}/R$, i.e., $\delta \phi =H/(2\pi )^{[9]}$, which is often
used in the analysis of the fluctuation spectrum of the inflationary
universe. We shall made an important supposition which says that the
relation $\delta \phi =H/(2\pi )$ should be suitable for all cases of
expanding universe, not merely for de-Sitter one. We can deem that this
assumption is reasonable due to the following four reasons. First, when we
apply the relation $\delta \phi =H/(2\pi )$ in the case of the early
inflation universe, the universe in this time is actually not an exact de
Sitter one, but we can use it successfully. Second, for each moment the
expanding universe can be taken approximately as a de-Sitter universe with
the expansion rate $H$. Third, the age of \ the universe is inverse
proportional to the Hubble constant $H$, the uncertainty of the energy is
proportional to $H$, the relation $\delta \phi =H/(2\pi )$ seems to coincide
with the uncertainty principle of the quantum mechanism. Four, in according
to the holographic hypothesis of the universe$^{[10]}$, some property of the
bulk of the universe is determined by its boundary, i.e., the horizon $H$.

In spite of $<\phi >$ is much larger than $\delta \phi $ during the period
of the inflation due to slowly rolling, however in contrast, now the quantum
fluctuation $\delta \phi $ which is not controlled by rolling process will
be much larger than the value $<\phi >$ for the present universe. Due to
nonzero value of $\delta \phi =H/(2\pi )$, the present universe has a
nonzero cosmological constant $V(H/(2\pi ))$. This is just our new
understanding to the essence of the actual nonzero cosmological constant,
i.e., it is a relic energy of the inflation potential.

\section{3.An unusual effective potential of the inflaton field}

{\hspace*{5mm} Due to the complications of generating the effective
inflation potential, we do not know its exact form at present. These
complexities include that the high loop quantum modifications, a running of
the parameters, the compactification and evolution of the internal
dimensions, and the various non-perturbative effects from the quantum field
theory, quantum gravity, superstring and even the M-theory. No matter how
complicated of its generating, we can always adopt an effective potential to
describe it, even though which form looks like very strange. }

We take the following effective potential for whole evolution of the
universe: 
\[
V(\phi )=m_{(\alpha )}^{4-\alpha }\left| \phi \right| ^{\alpha }+m_{(\beta
)}^{4-\beta }\left| \phi \right| ^{\beta }, 
\]
where the first term is responsible for the cosmological constant, the
second term is responsible for the chaotic inflation. We suppose that the
parameters $0<\alpha \lesssim 2/3$, $\beta \simeq 1.9$ and $m_{(\alpha )}\ll
m_{(\beta )}$, the reason will be given later on. We shall demonstrate that
this potential satisfies almost all requirements of the cosmological
phenomenology. The lowest point of this potential is indeed $V(0)=0$,
satisfying the naturalness condition suggested by us.

\section{4.Inflation and preheating}

{\hspace*{5mm} In according to the inflation theory, the slowly rolling
parameters are 
\[
\varepsilon =\frac{M_{p}^{2}}{2}(\frac{V^{\prime }}{V})^{2}\simeq \frac{%
\gamma ^{2}M_{p}^{2}}{2\phi ^{2}},\qquad \eta =M_{p}^{2}\frac{V^{\prime
\prime }}{V}\simeq \frac{\gamma (\gamma -1)M_{p}^{2}}{\phi ^{2}}, 
\]
where the parameter $\gamma =\alpha $ for $\phi \ll m_{(\alpha )}$ and $%
\gamma =\beta $ for $\phi \gg m_{(\beta )}$. The inflation should end at $%
\phi _{{\rm {end}}}\simeq 0.3M_{p}$ where $\varepsilon _{{\rm {end}}}-\eta _{%
{\rm {end}}}=\beta (2-\beta )\phi _{{\rm {end}}}^{-2}M_{p}^{2}/2$ $\simeq 1$%
. The parameter $\beta $ is not equal to $2$ in order to avoid $\varepsilon
-\eta $ being zero, otherwise the universe can not end its inflation. The
e-fold of the inflation is 
\[
N=\int \frac{Vd\phi }{M_{p}^{2}V^{\prime }}\simeq \frac{\phi _{{\rm {begin}}%
}^{2}-\phi _{{\rm {end}}}^{2}}{2\beta M_{p}^{2}}. 
\]
In order for our universe to have an enough inflating e-fold $e^{N}\gtrsim
10^{61}$ , our universe must begin to inflation at $\phi _{{\rm {begin}}%
}\gtrsim 23M_{p}$. Before $\phi _{{\rm {begin}}}$, the universe undergoes
the quantum production$^{[11]}$ and the chaotic inflation$^{[12]}$. The
density fluctuation of the universe is given by$^{[13]}$%
\[
\delta ^{2}=\frac{V}{150\pi ^{2}M_{p}^{4}\varepsilon }\simeq \frac{m_{(\beta
)}^{4-\beta }\phi ^{\beta +2}}{75\pi ^{2}M_{p}^{6}\beta ^{2}}, 
\]
which must be about an order magnitude of $10^{-10}$, then we obtain $%
m_{(1.9)}\simeq 2.2\times 10^{-6}M_{p}$ $=5.3\times 10^{12}$GeV, which is
far below the grand unification energy scale. The spectrum index $%
n=-6\varepsilon +2\eta \simeq $ $0.014$, the spectrum is almost scale
invariant. }

After the end of the slowly rolling, the inflaton field begins its fast
oscillation, which effective mass is $m_{\phi }^{2}=V^{\prime \prime }\simeq 
$ $\beta (\beta -1)m_{(\beta )}^{4-\beta }\phi ^{\beta -2}$. In the moment
of inflation end, the effective mass of inflaton is $m_{\phi {\rm {\ end}}%
}\simeq 0.7m_{(1.9)}$. The unstable vacuum energy $V_{{\rm {end}}%
}=m_{(1.9)}^{2.1}(\phi _{{\rm {end}}})^{1.9}\simeq $ $(6\times
10^{-4}M_{p})^{4}$ transforms into a lot of the zero temperature inflaton
particles. When the inflaton field rolls down, the field average value of
inflaton reduces, however the effective mass of the inflaton increases (if $%
\beta <2$) and becomes larger than the mass of fermion, which has a coupling 
$h\bar{\psi}\psi \phi $ with the inflaton, then all inflatons begin to decay
into pairs of the fermions. This turning point is determined by equation $%
h^{2}\phi _{{\rm {turn}}}^{2}=\beta (\beta -1)m_{(\beta )}^{4-\beta }\phi _{%
{\rm {turn}}}^{\beta -2}$. In this case $m_{\phi {\rm {\ turn}}}\simeq
1.3m_{\left( 1.9\right) }$ if $\ $the Yukawa coupling constant is $h\simeq 1$%
. The decay width is $\Gamma =h^{2}m_{\left( \beta \right) }/(8\pi )\simeq
2.7\times 10^{11}$GeV, which can be viewed as the preheating temperature $T_{%
{\rm {preh}}}$. \ This $T_{{\rm {preh}}}$ is not only enough low to avoid to
produce harmful topological detects, but also enough high to create some
super heavy long lifetime particles, which can just generate the highest
energy cosmic rays in our universe. If $\beta >2$, the inflaton will decay
too early to avoid producing dangerous topological defects. The fermion come
from the decay of the inflaton may be the gaugino, which can easily produce
all elementary particles of the supersymmtric standard model in a thermal
universe.

\section{5.The cosmological constant as a residual inflationary energy}

{\hspace*{5mm} When the inflaton field continue to roll down rapidly, the
part of the average value $<\phi >$ disappears, and bequeaths the part of
the quantum fluctuation $\delta \phi =H/(2\pi )$ as its value, which is very
small. Therefore the potential leaves a nonzero cosmological constant $%
\Lambda _{0}$, 
\[
m_{(\alpha )}^{4-\alpha }(\frac{H_{0}}{2\pi })^{\alpha
}=3M_{p}^{2}H_{0}^{2}\Omega _{\Lambda }, 
\]
we must have $m_{(\alpha )}=(3(2\pi )^{\alpha }\Omega _{\Lambda
}H_{0}^{2-\alpha })^{1/(4-\alpha )}M_{p}$, which magnitude order depends on
the index $\alpha $ sensitively. We list some data in following: $%
m_{(1)}\simeq 46$MeV with $w=-1/2$, $m_{(2/3)}\simeq $ $3.4$keV with $w=-2/3$%
, $m_{(1/2)}\simeq 55$eV with $w=-3/4$, where $w\equiv p/\rho =\alpha /2-1$
is a ratio of pressure over density, this formula will given later on. If $%
\alpha =0$, we have $m_{(0)}\simeq 2$meV with $w=-1$, i.e., true
cosmological constant, however in this case $V(0)\neq 0$ which does not
satisfy our requirement. This show that the cosmological constant is a relic
energy of the inflation potential. The parameter $\beta >1.8$ will guarantee
that the second term is far smaller than the first term of the inflaton
potential when the inflaton field value $\delta \phi $ is taken as $H/(2\pi
) $. }

During the evolution of our universe, the Friedmann equation is$^{[14]}$ 
\[
\rho _{m}+\Lambda =3M_{p}^{2}H^{2}. 
\]
Considering the fluctuation energy of the inflaton field introduced by us,
the above equation becomes 
\[
\frac{\rho _{0}}{R^{3}}+\frac{m_{\alpha }^{4-\alpha }}{(2\pi )^{\alpha }}%
\left( \frac{\dot{R}}{R}\right) ^{\alpha }=3M_{p}^{2}\left( \frac{\dot{R}}{R}%
\right) ^{2}. 
\]
This will affect the formation of the large scale structure of our universe
and the relevant observation data can be used to test our model! At a period
which redshift $z$ is between about $100$ and $10$, the inflaton energy term
can be omitted, the solution is approximately $R\propto t^{2/3}$, the matter
density term is $\rho _{m}\propto t^{-2}$, and the inflaton energy term is $%
V_{{\rm {vac}}}\propto t^{-\alpha }$, which is equivalent to the
quintessence with $w=\alpha /2-1$. The observation allow that $w<-2/3$,
therefore requiring to $\alpha <2/3$. If the observing data permits $w\simeq
-1/3^{[15]}$, we can combine the two term of our potential into a single
term with $\alpha =\beta =4/3$, i.e., the inflation and the cosmological
constant are uniformly described by only one term.

Actually our $V_{{\rm {vac}}}$ is not the quintessence by its self meaning.
As the universe continue to evolution, the difference between both becomes
very distinct. When time goes to infinite, $\Omega _{\Lambda }(\infty
)\rightarrow 1.0$. The vacuum energy will reduce, and the Hubble constant
will arrive a limit value 
\[
H_{\infty }=(\frac{\Omega _{\Lambda }(t_{0})}{\Omega _{\Lambda }(\infty )}%
)^{1/(2-\alpha )}H_{0}, 
\]
however if the cosmological constant is a true constant (no variable with
time, $\alpha =0$), the limited values $H_{\infty }$ will increase.

\section{6.A simple remark}

{\hspace*{5mm} In conclusion, we see that a simple potential model including
four parameters: two energy scales ($m_{(\alpha )}$, $m_{(\beta )}$) and two
power indices ($\alpha $, $\beta $), can explain almost all cosmological
phenomenology, from the inflation, slowly rolling, an enough e-fold, a
suitable density fluctuation, the flat spectrum index, preheating, avoiding
topological defects, explain the highest energy cosmic ray, to a nonzero
cosmological constant in spite of the true one being zero, and its special
evolution and affecting on the formation of the large scale structure in our
universe. Of course this model has an ability to include some cool dark
matters, such as WIMP. However, we face still on an intractable problem why
the universe chooses so strange inflation potential? How to calculate the
compactification of the extra-dimensions and the non-perturbation quantum
effects? }

\bigskip

{\bf Acknowledgment:}

{\hspace*{5mm} This work is supported by The foundation of National Nature
Science of China, No.19675038 and No.19777103. The author would like to
thank useful discussions with Prof. J.R.Bond, P.G.Martin, X.-M.Zhang,
X.-H.Meng and U.-L.Pen. }

\bigskip

{\bf References:}

[1]S.Perlmutter et al, Nature 391(1998)51.

[2]A.Riess et al, Astron.J.116(1998)1009.

[3]S.Weinberg, Rev.Mod.Phys.61(1989)61.

[4]E.Witten, Mod.Phys.Lett.A10(1995)2153.

[5]M.Turner, astro-ph/9904099.

[6]K.Freese, Nucl.Phys.B287(1987)797.

[7]D.N.Spergel and U.-L.Pen, ApJL491(1997)L67.

[8]M.Kamionkowski and N.Toumbas, Phys.Rev.Lett.77(1996)587.

[9]G.Gibbons and S.Hawking, Phys.Rev.D15(1977)2752.

[10]L.Susskind, J.Math.Phys.36(1994)6377.

[11]A.Vilenkin, Phys.Rev.D37(1988)888.

[12]A.Linde, \TEXTsymbol{<}\TEXTsymbol{<}Particle Physics and Inflationary
Cosmology\TEXTsymbol{>}\TEXTsymbol{>},

\qquad 1990 by Harwood Academic Publishers.

[13]H.Lyth and A.Riotto, hep-ph/9807278, to be in Phys.Rep.

[14]E.Kolb and M.Turner, \TEXTsymbol{<}\TEXTsymbol{<}The Early Universe%
\TEXTsymbol{>}\TEXTsymbol{>},

\qquad 1990 by Addison-Wesley Publishing Company.

[15]M.Turner and M.White, Phys.Rev.D56(1997)R4439.

\end{document}